\documentclass[aps,prl,showpacs,twocolumn,floats,epsfig]{revtex4}
\usepackage{amssymb}
\usepackage{amsbsy}
\usepackage{amsmath}
\usepackage{epsfig}

\begin{document}
\def\be{\begin{equation}}
\def\ee{\end{equation}}
\def\bearr{\begin{eqnarray}}
\def\eearr{\end{eqnarray}}
\def\tc{$T_c~$}
\def\tcl{$T_c^{1*}~$}
\def\c2{ CuO$_2~$}
\def\ruo{ RuO$_2~$}
\def\lsco{LSCO~}
\def\bi{bI-2201~}
\def\tl{Tl-2201~}
\def\hg{Hg-1201~}
\def\sro{$Sr_2 Ru O_4$~}
\def\rc{$RuSr_2Gd Cu_2 O_8$~}
\def\mgb{$MgB_2$~}
\def\pz{$p_z$~}
\def\ppi{$p\pi$~}
\def\sqo{$S(q,\omega)$~}
\def\tperp{$t_{\perp}$~}
\def\he4{${\rm {}^4He}$~}
\def\ags{${\rm Ag_5 Pb_2O_6}$~}
\def\nxcob{$\rm{Na_x CoO_2.yH_2O}$~}
\def\lsco{$\rm{La_{2-x}Sr_xCuO_4}$~}
\def\lco{$\rm{La_2CuO_4}$~}
\def\lbco{$\rm{La_{2-x}Ba_x CuO_4}$~}
\def\half{$\frac{1}{2}$~}
\def\tst{${\rm T^*$~}}
\def\tch{${\rm T_{ch}$~}}
\def\jeff{${\rm J_{eff}$~}}
\def\nbc{${\rm LuNi_2B_2C}$~}
\def\cabc{${\rm CaB_2C_2}$~}
\def\nboo{${\rm NbO_2}$~}
\def\voo{${\rm VO_2}$~}
\def\h2o{${\rm H_2 O}$~}
\def\nh3{${\rm N H_3}$~}

\title{Silicene and Germanene as prospective playgrounds for\\
Room Temperature Superconductivity}

\author{G. Baskaran\\
The Institute of Mathematical Sciences, Chennai 600 113, India\\
Perimeter Institute for Theoretical Physics, Waterloo, Ontario, Canada N2L 2Y5}

\begin{abstract}
Combining theory and certain striking phenomenology we suggest that silicene and germanene are \textit{elemental Mott insulators} and abode of doping induced high Tc superconductivity. In our theory, a 3 fold reduction in silicene $\pi$-$\pi^*$ band width, in comparison to graphene, and short range coulomb interactions enable Mott localization. Recent experimental results are invoked to provide support for our Mott insulator model: i) a significant $\pi$-band narrowing, in silicene on ZrB$_2$ seen in ARPES, ii) a superconducting gap appearing below 35 K with a large $\frac{2\Delta}{k_BTc}\sim$ 20 in silicene on Ag, iii) emergence of electron like pockets at M points, on electron doping by Na adsorbent, iv) certain coherent quantum oscillation like features exhibited by silicene transistor at room temperatures and v) absence of Landau level splitting upto 7 Tesla and vi) superstructures, not common in graphene but, ubiquitous in silicene. A synthesis of the above results using theory of Mott insulator, with and without doping, is attempted. We surmise that if competing orders are taken care of and optimal doping achieved, superconductivity in silicene and germanene could reach room temperature scales; our estimates of model parameters, t and J $\sim$ 1 eV, are encouragingly high, compared to cuprates.
\end{abstract}

\maketitle
\section{Introduction}

A wealth of activity in the field of graphene, following the seminal work of Novosolev and Geim \cite{NovosolevGeim1,NovosolevGeim2,NovosolevGeim3}, has paved way for silicene \cite{slcnPrediction1,slcnPrediction2,slcnPrediction3,slcnReview}, a silicon analogue of graphene. This new entrant might have a potential to begin another fertile direction in condensed matter science and technology. Replicating a rich graphene physics has been a part of recent efforts. Stable silicene layer has been created on a few metallic substrates, Ag, ZrB$_2$ and Ir. However, synthesis of free standing silicene remains a challenge. Interesting ARPES and STM results are available \cite{slcnAg1,Vogt,slcnAg2,slcnZrB2,slcnIr,ChenSTS,slcnLandauLevel,NaDoping}. Silicene based field effect transistor has been also fabricated \cite{Akinwande1}. There have been successful attempts to synthesize germanene and related systems \cite{germanene} on certain metallic substrates.

Aim of the present article is to provide a low energy model, a rather unexpected one, for electrical and magnetic properties of silicene. Finding a suitable low energy model for strongly interacting quantum matter continues to be a challenge. The very experimental results we wish to understand guide us to correct theoretical modelling. Theory in turn guides experiments. The synergy continues. This is true from Standard Model building in elementary particle physics to Standard Model building for cuprate. Silicene is no exception.

The currently prevalent view is that neutral silicene is a Dirac Metal, a semimetal qualitatively similar to graphene \cite{slcnReview}. Purpose of the present paper is to offer a different view point, that \textit{Silicene is not a Carbon Copy of Graphene} - \textit{it is an elemental Mott insulator}.  If proved correct, our provocative proposal will make silicene different from semi metallic graphene in a fundamental fashion and open new avenues for physics and technology, arising from strong electron correlation effects. 

It is well appreciated now, thanks to  the path breaking discovery of high Tc superconducting cuprates by Bednorz and Muller \cite{BednorzMuller} and subsequent resonating valence bond theory by Anderson and collaborators  \cite{PWAScience,BZA,BAGauge,GBIran,WenBook}, that Mott insulators and strong electron correlations are seats of a variety of rich physics and phenomena. In addition to superconductivity it includes, quantum spin liquids, emergent fermions with Fermi surfaces, gauge fields, quantum order, topological order and so on. Further, inspired by certain recent theoretical development \cite{muramatsuQSL} there is a debate and search for quantum spin liquids in honeycomb lattice Hubbard model \cite{sorellaTosatti,sorellaQSL,hassanQSL}. We believe that silicene and germanene will fit well into the discussion as real candidate materials, albeit with added novel features.

Several ab-initio calculations\cite{slcnPrediction1,slcnPrediction2,slcnPrediction3}, many body theory \cite{KatznelsonSlcn}, quantum chemical calculations and insights \cite{Sheka1,Sheka2,March1,March2,Hoffman,Ayan} are available for silicene. Interestingly there is a differing view, which has not been well appreciated. Existing solid state manybody calculations predict stable free standing semi metallic silicene that is a similar to graphene. However, quantum chemical methods and insights doubt existence of a stable free standing silicene \cite{Sheka1,Sheka2,Hoffman} because of radicalization/reactivity and reduced aromaticity, arising from a weakened p-$\pi$ bond. 

It is clear that theory of silicene is challenging and less understood compared to graphene, because of a growing importance of electron electron interaction and a soft c-axis displacement (puckering) degree of freedom, arising from an easy sp$^3$ mixing. Our model and theory is aimed to initiate new discussion, theoretical and experimental studies.

The present article is organized as follows. We interpret certain existing theoretical results, Coulomb screening argument for Mott transition and quantum chemical insights as providing support for our proposal of a narrow gap Mott insulating state for neutral silicene. We also identify and discuss a set of about 6 anomalous experimental results that point to a Mott insulating state. 

A Heisenberg model, containing additional multispin interactions, is introduced to describe spin dynamics in a small gap Mott insulator, and a tJ model for the spin-charge dynamics of doped Mott insulator. Then we discuss the aforementioned anomalous experimental results in the light of our model.

Prospects for high Tc superconductivity, within our model, is discussed next. In view of larger and more favourable tJ parameters, in comparison to layered cuprates, there is a distinct possibility of \textit{a Tc approaching room temperature scale} provided competing interactions are taken care of. Using existing theoretical works we come to the conclusion that superconductivity is likely to be a chiral spin singlet d + id superconductivity. 

Superstructures, not common in graphene but ubiquitous in silicene grown on substrates, are interpreted as arising from a strong response of c-axis deformable Mott localized p-$\pi$ electrons, to substrate perturbations, through site dependent sp$^3$ mixing. In a recent ab-initio calculation with Vidya \cite{VidyaGB} we have we have found indirect evidence for Mott localization, through presence of sizable Kekule (valence bond) order and a weak antiferromagnetism. Further we have interpreted experimentally seen reconstructions as Kekule or Valence Bond order.

The present paper assumes that tJ model in 2d describes Kosterlitz-Thouless superconductivity for a range of doping. While there is no rigorous proof for this, it is well certified by a body of analytical, numerical and experimental efforts available for the square lattice cuprates, ever since Bednorz Mullers discovery and the beginning of RVB theory. Existing studies on honeycomb lattice tJ model indicate that spin singlet superconductivity continues to be a dominating phase over a range of doping, but with d + id order parameter symmetry.

Our present work is a natural extension of our earlier work on graphene \cite{GBMgB2,GBAkbar,GBPathakShenoy}. It has been our view that graphite and graphene should show interesting electron correlation effects, even though electron-electron interaction strengths are moderate compared to the band width. According to us, \textit{reduced 2-dimensionality of graphene causes an amplification of electron correlation effects}. Our earlier prediction of spin-1 collective mode in neutral graphene \cite{GBAkbar} and very high Tc superconductivity in doped graphene \cite{GBMgB2,GBPathakShenoy} are based on use of a moderate electron repulsion strength. Our main message in the present article is that 2d silicene and germanene, having a third of graphene band width, but two thirds of on site coulomb interaction strength of graphene should exhibit more pronounced electron correlation effects. 

In a recent article we have suggested a \textit{Five fold way} to new high Tc superconductors \cite{5FoldWayGB} one of the ways is \textit{graphene route}. Silicene and germanene are most likely to lie in the graphene route and help us achieve room temperature superconductivity.

\section{Is Free Standing Silicene Stable ?}

Beginning with hexasilabenzene, Si$_6$H$_6$, a silicon analogue of benzene, chemists have wondered about the existence of stable planar p-$\pi$ bonded Si based molecules \cite{SiBenzene1,March1}. Their concern is that an increased Si-bond length, a simple consequence of a 60 \% increase in atomic radius, will weaken the p$-\pi$ bond, leading to reduced aromaticity and increased chemically reactivity. It is also an experimental fact that free Si$_6$H$_6$ molecule has not been synthesized so far. 

Silicene is an infinitely extended planar p-$\pi$ bonded network of Si atoms. 
DFT calculations and beyond argue for a stable free standing graphene. On the other hand, Sheka in 2009 \cite{Sheka1,Sheka2} and Hoffman \cite{Hoffman} in 2013, have questioned stability and very existence of free standing silicene because of reduced aromaticity and enhanced radicalization. This, according to them, will make silicene react with any molecular dirt.

Radicalization in quantum chemistry is creation of unpaired lone electron in certain molecular orbitals in the ground state; the loners are generically weakly coupled to other loners, if present. In the context of periodic systems such as a crystalline solid we interpret it to mean an extreme Mott localization and formation of nearly decoupled spins. In this Mott insulating state quantum fluctuations lead to residual (superexchange) couplings among spins quantum magnetism.

Free standing silicene has not synthesized experimentally so far. A key stabilizing factor, namely metallic substrates, Ag, Ir and ZrB$_2$ is needed. Further, a strong hybridization between Ag bands and Si orbitals at the fermi level leads to significant modification of electronic properties of free standing silicene.

In what follows, by \textit{stable Mott insulating single layer silicene} we mean the following. We assume that (metallic) substrate stabilizes a single layer silicene and at the same time does not significantly modify the Mott or doped Mott insulator character, that we are after.

 %\begin{widetext}
\begin{figure*}[t]
\epsfxsize 12cm
%\centerpage
%{\epsfbox{FleurencePRLFig3.pdf}}
\includegraphics{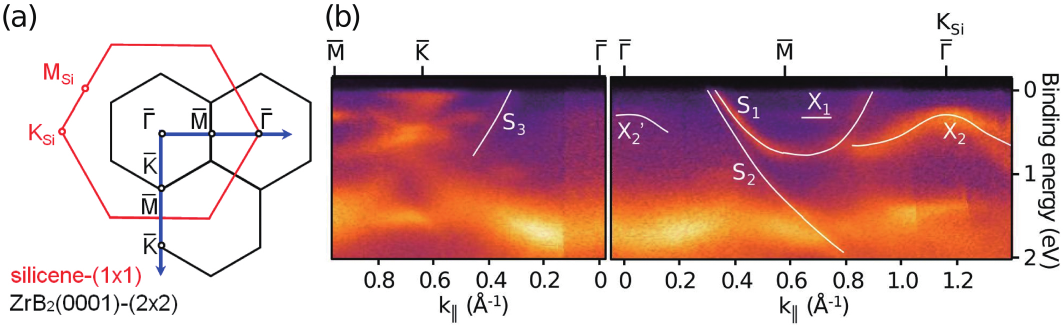}
%\end{widetext}
%\begin{widetext}
\caption{Figure 1. ARPES data of silicene on ZrB$_2$ (reproduced from reference \cite{slcnZrB2} with permission). 
A hole like narrow band of width (marked X$_2$), below a gap of 0.3 eV at K point is seen in figure 1a . This band is only 1 eV wide, compared to 6 eV $\pi-\pi^*$ band width given by LDA. It has appreciable spectral weight only in one third of the Brilluoin zone. We interpret this band as a strongly renormalized hole band of a Mott insulator. Interestingly all silicene bands X$_1$, X$_2$ and X$_2$' lie within an energy interval of 0.5 eV. S's refer to substrate bands.}
\end{figure*}
%\end{widetext}
\section{Mott Insulating Silicene - Theoretical Support} 
In this section, using recent estimates of Hubbard U, nearest neighbour repulsion V and Mott's argument for metal insulator transition invoking screened long range coulomb, we will argue a Mott insulating ground state for silicene.

We begin with a summary of basic quantum chemistry and band theory results for silicene. Silicon, located just below carbon in the periodic table has a larger atomic radius $\sim$ 1.17 $\AA$A, in contrast to a smaller value $\sim$ 0.77 $\AA$ for carbon. This leads to a \textit{significant} 3p-$\pi$ \textit{bond stretching}: Si-Si distance is $\sim$ 2.3 $\AA$, while C-C distance is only 1.4 $\AA$ in graphene. 

Unlike graphene, Si allows a small sp$^3$ admixture to sp$^2$ bonding, resulting in puckered sigma bonds. That is, Si atoms of the two triangular sublattices of the honeycomb net undergo a small and opposite displacements, leading to a net c-axis stretching $\sim ~0.4$ Au. A reduced 3s-3p level separation in Si atom, compared to 2s-2p level separation in C atom encourages a small sp$^3$ hybridization.

Electronic structure calculations \cite{slcnPrediction1,slcnPrediction2,slcnPrediction3} predict a semimetal band structure qualitatively similar to graphene, containing two Dirac cones at K and K' points. A major differences from graphene is a 3 fold reduction in the $\pi-\pi^*$ band width.

Following a pioneering work of Sorella and Tosatti \cite{sorellaTosatti} a recent estimate \cite{muramatsuQSL} gives an accurate value of the critical value $\frac{U}{t} \sim$ 3.8, for metal to Mott transition in Hubbard model on Honeycomb lattice. In a very recent many body theory Schuler et al. \cite{KatznelsonSlcn} estimate a value $\frac{U}{t} \sim$ 4.1 for silicene. This puts silicene on the Mott insulator side, close to the phase transition point. However, Schuler et al. argue that inclusion of nearest neighbour repulsion $\frac{V}{t} \sim $ 2.31, will reduce U to an effective $\tilde{U} \approx$ U - V and bring silicene back to a semi metallic state like grapheme. In what follows we argue, on the contrary, that presence of finite V will hasten Mott localization and reinforce a Mott state through a first order phase transition.

We start with Mott's argument for a first order Mott transition. Mott begins by asking whether a \textit{screened coulomb interaction} present in the metallic state is sufficient to form a quantum mechanical \textit{bound state of an electron and the hole it left behind at its home site}, at the fermi level. Within a Hubbard model idealization this happens when band width becomes comparable to onsite U; it also means that at very large U every site binds a lone daughter electron. In Mott's argument the transition is pre-empted by reduced screening of long range interaction (as we loose free carriers by bound state formation), through a feed back resulting in a first order phase transition. This aspect is not contained in the simple Hubbard model. 

Let us assume that in addition to U we have a non-zero nearest neighbour coulomb interaction V, which is a leading term in the long range part of screened coulomb interaction. According to Mott's arguments, V will add to the already existing on site attraction U between an electron and the hole it left behind. This additional attraction (widening of the potential well) decreases kinetic energy of bound state. Resulting increase in binding energy hasten bound state formation. Roughly, the onsite Hubbard U gets enhanced: $\tilde{U} \approx$ U + $\alpha $V, where $\alpha \sim$ 1. 

Thus we conclude, by what we believe to be a correct use of important estimates of Schuler et al. \cite{KatznelsonSlcn} that silicene is in the Mott insulating side of the metal insulator transition point. We hope to address this point in some detail in a separate paper.

\section{Mott Insulating Silicene - Phenomenological Support}

In this section we briefly review experimental results in silicene which are anomalous, from the point of view of a Dirac metal. But, as we will discuss latter, they seem normal from Mott insulator point of view. 

$\bf Doped ~ Hole~and~a~Narrow~Band$: In a work that has received a wide attention, using ARPES, Vogt et al. \cite{Vogt} show presence of a Dirac cone dispersion over a wide energy range below Fermi level, with a v$_F$ comparable to that in graphene. There is an ongoing debate \cite{VogtDebate1,VogtDebate2,VogtDebate3} on whether this is  primarily a silicene p-$\pi$ band or primarily a Ag metallic bands or a strong hybrid. Recent ARPES result shows \cite{KuheinBansilARPES} multiple (more than the expected double) Dirac cones in the silicene on Ag system. This has been attributed to a strong hybridization of silicene and silver states at the fermi level.

On the other hand, ARPES study of silicene grown on ZrB$_2$ by Fleurence et al. \cite{slcnZrB2} shows a
very different behaviour. ZrB$_2$ is a low carrier density metal with small Fermi pockets. It is expected to have less electronic influence on silicene. Fleurence et al. find a remarkably narrow band around around K and K' points (marked X$_2$ in figure 1b), lying below a finite gap $\sim$ 0.3 eV. Its spectral weight vanishes in two-third of the Brillouin zone. An extrapolation from the shape of the visible part of the band gives us a total band width $\sim$ 1 eV. This is to be contrasted with the total $\pi-\pi^*$ band width of 6 eV as given by LDA calculations. Thus there is a band width renormalization by a factor of 10. Further 3 silicene bands denoted by X$_1$, X$_2$ and X$_2$' all lie in an energy interval of 0.5 eV. This spectrum is reminiscent of hole spectrum in Mott insulating \lco.

$\bf A~Superconducting~Gap~Anomaly$:  Chen et al. have observed \cite{ChenSTS} a superconductor like gap structure in their STS study of silicene on Ag and a Tc $\sim$ 35 to 40 K. They provide arguments in support of a gap arising from superconductivity. A large $\frac{2\Delta}{kT_c} \sim 20 $ makes it anomalous.
Further experiments are needed to substantiate this important result.  

$\bf Absence~of~Landau~Level~formation$: An STM work \cite{slcnLandauLevel} in the presence of a magnetic field as large as 7 Tesla, do not find an expected Landau level splitting. In our view, a doped hole looses nearly all its quasiparticle weight because of the strong electron-electron interaction. We find that quasiparticle line broadening, as inferred from experiment is larger than the expected Landau level splitting, making it invisible in STS measurements. In graphene, because of quasi particle coherence, one sees Landau level structures for similar magnetic fields in the STS measurements. 

$\bf Silicene~Lattice~Reconstructions:$ Recent experiments on silicene grown on metal surfaces exhibit \cite{VB1,VB2,VB3} lattice reconstructions, $\sqrt{3}\times \sqrt{3}$, $\sqrt{7}\times \sqrt{7}$,  $\sqrt{13}\times \sqrt{13}$, 4 $\times$ 4 etc. They are temperature dependent and some of them exhibit finite temperature phase transitions. Further, reconstructions are accompanied by \textit{space dependent sp$^3$ mixing and consequent bond length modulation}.  Even within band theory there is no Fermi surface. So a standard route for density wave instabilities namely Fermi surface nesting is absent. This explains why reconstructions or CDW order is not seen in graphene. Graphene is stiff, inview of a $\sim$ 9 eV wide filled $\pi$-band and an associated large \textit{aromaticity}. We suggest that p-$\pi$ electrons in silicene, because of Mott localization, respond strongly to substrate perturbations by making use of the soft sp$^3$ hybridization option and corresponding c-axis displacements. 

$\bf Fermi~Arc~like~Electron~Pockets$: Electron doping in silicene deposited on ZnB$_2$ by alkali metal leads to appearance of small Fermi arc or pockets like features at the M points \cite{NaDoping}. This is not easily explained within LDA band structure result. A doped Mott insulator on the other hand, does not obey Luttinger theorem in the usual fashion and can have unusual emergent Fermi pockets and Fermi arcs - a striking example is the under doped cuprates.

\textbf{Silicene Transistor and Resistance Oscillations} Using an innovative technique a silicene transistor has been successfully fabricated \cite{Akinwande1}. It exhibits a desired on off resistance radio as gate voltage is varied. Further, it exhibits an intriguing resistance oscillation at room temperatures, resembling a Fabry-Perot type of quantum interference.

Thus we have some theoretical arguments and a few phenomenological facts in support of a Mott insulator picture for neutral silicene. In the following we will build a model, address the above experimental observations from the Mott insulator point of view. 

\section{Spin Liquid State in Silicene ?}

Let us first discuss nature of the hypothesized Mott insulating state of neutral silicene. From ARPES experiment \cite{slcnZrB2} we estimate a Mott Hubbard gap of $\sim$ 0.6 eV. In transition metal oxide and organic Mott insulators, Mott gap is often comparable to band widths. As the inferred Mott gap, 0,6 eV is only a tenth of the total $\pi-\pi^*$ band width $\sim$ 6 eV silicene is \textit{a Mott insulator with a small charge gap}.

In a Mott insulator low energy degree of freedom are spins and spin-spin interaction arise from super (kinetic) exchange processes. One can estimate J, using the standard expression, J $ \approx \frac{4t^2}{U^2}$. The values, t $\sim$ 1.14 eV and an effective  U $\sim$ 5 eV, discussed earlier for silicene,  gives us a J $\approx$ 1 eV.
As J is comparable to the Mott Hubbard gap of 0.6 eV, a strong virtual charge density and charge current fluctuations will renormalize J to lower values. Further, higher order cyclic exchanges will be also present.

Ignoring spin-orbit coupling for the moment, our effective spin Hamiltonian in the Mott insulating state is a spin-half Heisenberg Hamiltonian
\be
H_s = J \sum_{\langle ij \rangle} ({\bf S}_i \cdot {\bf S}_j - \frac{1}{4}) + {\rm 4~\&~6~spin~terms}
\ee
If the nearest neighbour J dominates we will have long range antiferromagnetic order in the ground state. 
As recent results show, within nearest neighbour repulsive Hubbard model one is unlikely to stabilize a spin liquid phase \cite{sorellaQSL}. 
In the present case, because of softness of puckering, there may be valence bond order, rather than a spin liquid, without doping. However, as in cuprates, this may not be an important practical issue. This is because we expect that even a small density of dopants through their dynamics will destroy long range AFM order and valence bond order and stabilize some kind of spin liquid state (pseudo gap phase) containing incoherent dopant charges. It is this doping induced spin liquid state that will determine nature of superconductivity over a range of doping. It is likely that a variety of spin liquids are around the corner.

\section{Dopped Mott Insulator}

For dopant dynamics in a Mott insulator, upto certain range of doping, a projective constraint arising from upper and lower Hubbard band formation and surviving superexchange are important. Thus the relevant model for doped Mott insulator is 
the tJ model:
\be
H_{\rm tJ} = - t \sum_{\langle ij \rangle} ( c^\dagger_{i\sigma}c^{}_{j\sigma}  + h.c.)  + J \sum_{\langle ij \rangle} ({\bf S}_i \cdot {\bf S}_j - \frac{1}{4}n_i n_j)
\ee
with the local constraint $ n_{i\uparrow} + n_{i\uparrow} \neq$ 2 or 0, for hole and electron doping respectively. The value of J, t  $\sim$ 1 eV.

It is known from the study of single hole dynamics in Mott insulating cuprates that a free and coherent propagation of a doped hole or electron (carrying its charge and spin) is frustrated by the strongly quantum entangled spin background. This leads to a significant band narrowing and loss of spectral weight over a large part of the Brillouin zone, as seen in ARPES experiments \cite{KyleShen} and tJ model calculations 
\cite{tJHole}. For cuprates band theory gives a width of 8t $\approx$ 3 to 4eV, while ARPES results in the Mott insulating cuprates give a band width of 2J $\sim$ 0.3 eV (essentially spin wave band width) for holes. This is a substantial, ten fold reduction of hole band width.

As we saw earlier ARPES study on silicene grown on ZrB$_2$ gives a parabolic hole band at K point, about 0.3 eV below the Fermi level (band X$_2$ in figure 1 b) with a band width $\sim$ 1 eV. Comparing results with cuprates mentioned above, we get a J $\sim$ 0.5 eV. This is in the right ball park, as our estimate of 
J $\sim$ 1 eV.

\section{Doped Mott Insulator and Superconductivity} 

We start with tJ model for our doped Mott insulator in a honeycomb lattice and discuss superconductivity.
Interestingly, the above model (equation 2) was studied in a mean field approach first in \cite{ChoyMcKinnon} for graphite. Later the present author independently studied the same model ignoring onsite constraints \cite{GBMgB2} as a semi microscopic way of incorporating Pauling's spin singlet (resonating  valence bond) correlations in planar \textit{graphitic systems}, within a band theory approach. It leads to a prediction of very high Tc superconductivity around an optimal doping. Our meanfield theory result of high Tc superconductivity for graphite was reanalyzed and confirmed by Annica Black-Shaefer and Doniach \cite{AnnicaDoniach}. They further found a remarkable spin singlet chiral superconducting state, namely a state with d + id symmetry as the most stable mean field solution for the same range of doping. A more conservative repulsive Hubbard model has been studied from superconductivity point of view using variational Monte Carlo method in reference \cite{GBPathakShenoy}, by us and collaborators for graphene. There are other important works addressing the same issue\cite{Honerkamp1,dungHai,Thomale,Levitov,Honerkamp2,dPlusIDBilayer} by different methods. They all support the possibility of d + id chiral spin singlet superconductivity. 

A very recent work \cite{tJGuWen} studies tJ model on a honeycomb lattice using a new and powerful variational approach using Grassman Tensor Network states. It confirms d + id superconductivity for tJ model on honeycomb lattice for a range of doping. 
\begin{figure}[t]
\epsfxsize 7cm
\centerline {\epsfbox{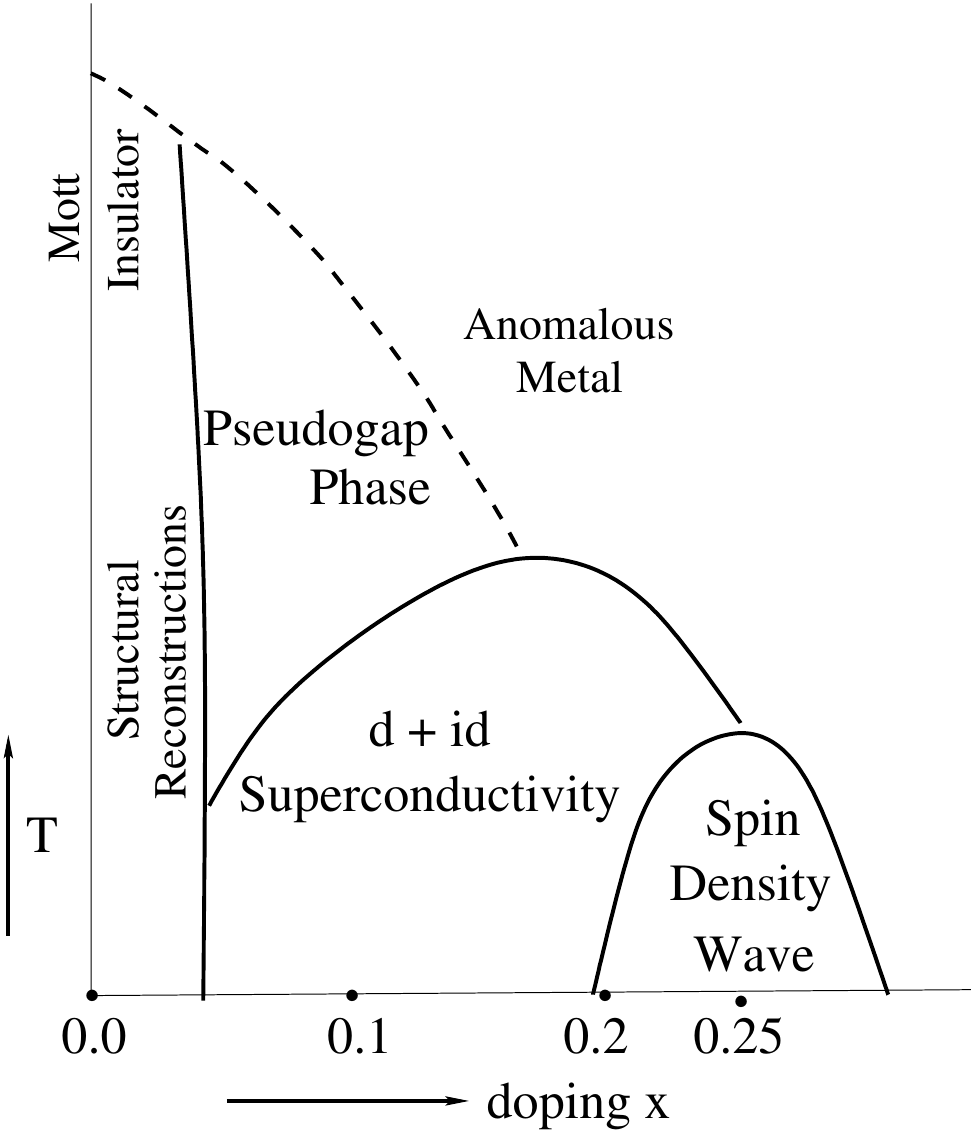}}
\caption{A schematic phase diagram for doped silicene and graphene.}
\end{figure}
Using available theory and insights we suggest a phase diagram (figure 1), qualitatively similar to cuprates, including the \textit{pseudo gap phase}. A new aspect for the honey comb lattice is presence of a van Hove singularity and an associated nesting (figure 1) at a doping of x = 0.25. It has been suggested that this nesting might stabilize a complex magnetic order \cite{TaoLiVanHove} or charge density wave order.  If our variational calculation \cite{GBPathakShenoy} performed for graphene, a doped semimetal described by intermediate repulsive U Hubbard model, is any guidance, maximum superconducting Tc occurs already around 15 percent, somewhat similar to cuprates. 

The issue of scale of superconducting Tc is very exciting and it shows some promise. As mentioned earlier we have an unusually large J $\sim$ 1 eV and t $\sim$ 1 eV for our honeycomb lattice. This is to be contrasted with a value of J $\sim$ 0.15 eV and t $\sim$ 0.25 eV for the square lattice cuprates, known for their record 
Tc $\sim$ 90$^{\circ}$ K, for for single layer cuprates. We have an average 4 fold increase in t and J for silicene.  Does silicene offer a 4 fold increase in Tc ? Taking care of lattice structure difference  between cuprate and silicene crudely, a scaling gives at least 3 fold increase of Tc. Thus room temperature scales for Tc seems within reach, provided competing orders are taken care of .

As in cuprates, we expect competing orders, charge and spin stripes to challenge the high Tc superconducting states. Further, as we will see, electron lattice interaction can cause patterns of vertical displacements of Si atoms, arising from sp$^3$ mixing, It is conceivable that there will be stripes and 2d patterns, corresponding to various valence bond orders. These valence bond localization tendencies will compete and reduce Tc significantly. 

The only available, but remarkable indicator for superconductivity in silicene is an STS study \cite{{ChenSTS}}. They find a Tc in the range 30 - 40 K and a large gap of 35 meV; that is, an anomalous value, $\frac{2\Delta}{k_B Tc} \approx 20 $. If we take the maximal gap value from this experiment and extract a mean field Tc, using weak coupling BCS result, we get a number close to 200 K ! We interpret this anomaly in $\frac{2\Delta}{k_B Tc}$, as due to a superconducting state with a large local pairing gap but with Tc reduced by a strong phase fluctuations arising from a static or low frequency local competing orders. This needs to be investigated theoretically and experimentally further. 

We have interpreted \cite{GBFabryPerot} a quantum oscillation in resistance seen in silicene transistor at room temperatures, as a Fabry-Perot type of interference of preformed bosonic charge 2e Cooper pairs, and physics similar to the pseudo gap phase of cuprates. Briefly, the gate induced carriers (holons or doublons) enable delocalization and interference of preformed charge 2e singlets, even before full phase coherence and superconductivity develops. 

There are recent suggestions of superconductivity in silicene based on different mechanisms \cite{KeihuTheory,phononMechanism}. We wish to point out that Liu et al. \cite{KeihuTheory} have theoretically studied possibility of spin fluctuation induced chiral d + id superconductivity in \textit{neutral} bilayer silicene. In their mechanism, intralayer hopping between the two Dirac metallic layers produce small electron and hole pockets around K and K' points. In our theory such a bilayer will remain insulating with an additional feature of a spin gap induced by a strong interlayer exchange coupling (spin singlet bond formation between Si atoms along c-axis). External doping or gate doping by very strong electric fields applied perpendicular to the layers will be needed to create superconductivity in the Mott insulating bilayers.

\section{Germanene and Stanene}

Group IV elements in periodic table C, Si, Ge, Sn and Pb have increasing atomic radii
0.77, 1.17, 1.22, 1.45 and 1.8 $\AA$ respectively. The p-$\pi$ bonds in silicene, germanene and stanene get weaker because of increasing atomic radii, in spite of increasing size of the p orbitals. Ab-initio calculations for germanene leads to a few percent enhancement of Ge-Ge bond length, but a substantial increase of puckering and c-axis stretching from $\sim$ 0.4 $\AA$ in silicene to $\sim$ 0.6 $\AA$ in germanene. There is a small reduction in $\pi$ and $\pi^*$ band widths from 3 to 2.5 eV. 

All things being quantitatively nearly equal at the level of model, germanene should be a narrow band Mott insulator. Consequences we have discussed so for, including possibility of quantum spin liquid and high Tc superconductivity on doping should be anticipated. There are interesting experimental developments with respect to germanene and stanene recently \cite{germanene}.

\section{Summary and Discussion}

To summarise, we have hypothesized that silicene and germanene are narrow gap Mott Insulators. This challenges the widely held belief that they are a Dirac metal, like graphene. In making our proposal we have relied heavily on a synthesis of  theory and phenomenology. We have used ARPES, STM and recent conductivity results. Mott's arguments, quantum chemical insights and extensive theoretical study of graphene and silicene has helped us in our proposal. Using known results we discussed various solutions for our model, including spectral function of a hole in a Mott insulator, absence of Landau level splitting, superconductivity and so on. Unconventional superconducting order namely d + id and high Tc's was also discussed. 

An important support for Mott localization comes from our translation of the quantum chemical results of Sheka \cite{Sheka1,Sheka2} to solid state terminology. Sheka studies nano silicene clusters and finds an extensive radicalization. Radicalization is extracted using a procedure that uses spin polarized (antiferromagnetic) ab-initio solutions. This interesting quantum chemical approximation is less known in solid state context. As indicated earlier we equate \textit{radicalization in finite systems to Mott localization in extended solid state context}. Thus Sheka's finding of extensive radicalization and alleged unstable silicene is a conservative evidence for narrow gap Mott insulator formation and strongly exchange coupled spins.

We have ignored spin-orbit coupling in the present paper, simply to focus on the key physics of strong correlations. Spin orbit coupling will play its own, important unique role, some what different from standard band insulator or semi metal. Within the context of semi metallic graphene interesting effects of spin orbit coupling are being studied \cite{Liu,Ezawa,Nagaosa}

Direct and indirect methods should be used to unravel an underlying Mott insulator. Optical conductivity, $\sigma(\omega)$ measurement is an urgent one. It should focus on finding signatures on Hubbard band features. It will be interesting to confirm the existing claim of superconducting gap  in STS measurements \cite{ChenSTS}. Reconstructions should be studied carefully to distinguish valence bond density wave and plaquette resonance density waves. Pseudo gap physics needs to be explored using NMR and NQR measurement. Our tantalizing prediction of very high Tc superconductivity reaching room temperature scales needs to be explored. This needs ways of understanding and overcoming unavoidable competing instabilities such as spin and charge stripes.

Next major experimental and theoretical challenge is to see whether a free standing silicene exists and if it can be synthesized and studied. 

I am not aware of any real material for which tJ model parameters are as large as we have suggested for silicene and germanene,t, J $\sim$ 1 eV. So we consider silicene and germanene as forefront materials in the race for room temperature superconductivity.

{\bf Acknowledgement} I thank Prof. Yamada-Takamura for giving permission to reproduce a figure from \cite{slcnZrB2}; Dr Ayan Datta for a discussion; Dr Kehui Wu and colleagues for informative discussions at the Silicene meet at Beijing in 2014. I thank Science and Engineering Research Board (SERB), Government of India for the SERB Distinguished Fellowship.  Additional support was provided by the Perimeter Institute for Theoretical Physics. Research at Perimeter Institute is supported by the Government of Canada through the Department of Innovation, Science and Economic Development Canada and by the Province of Ontario through the Ministry of Research, Innovation and Science.

%\begin{widetext}

%\end{widetext}
\end{document}